\documentclass[12pt,english]{article}
 \usepackage{helvet}
 \usepackage[T1]{fontenc}
 \usepackage[cp1251]{inputenc}
 \usepackage{geometry}
 \usepackage{graphicx}
 \usepackage{setspace}
 \usepackage{babel}

 \def\gsim{\lower.4ex\hbox{$\;\buildrel >\over{\scriptstyle\sim}\;$}}
 \def\lsim{\lower.4ex\hbox{$\;\buildrel <\over{\scriptstyle\sim}\;$}}
 \def\newpage{\vfill\eject}
 \def\bl{\par\vskip 12pt\noindent}
 \def\bll{\par\vskip 24pt\noindent}

 \def\beg{\begin{eqnarray}}
 \def\ende{\end{eqnarray}}

 \def\araa{ARA\&A}
 \def\aa{A\&A}
 \def\apj{ApJ}

 \def\mnras{MNRAS}
 
 \def\sp{Solar Phys.}
 \def\aj{Astron. Rep.}
 \def\paj{Astron. Lett.}

\begin{document}
\

\vskip 2.5 cm

\begin{center}
NORTH-SOUTH ASYMMETRY OF SOLAR DYNAMO \\[0.2truecm]
IN THE CURRENT ACTIVITY CYCLE
\end{center}

\bll

\centerline{L.\,L.~Kitchatinov$^{1,2}$\footnote{E-mail: kit@iszf.irk.ru}, A.\,I.~Khlystova$^1$}

\bl

\begin{center}
$^1${\it Institute for Solar-Terrestrial Physics, Lermontov Str. 126A, Irkutsk, 664033, Russia} \\[0.1 truecm]
$^2${\it Pulkovo Astronomical Observatory, Pulkovskoe Sh. 65, St. Petersburg, 196140, Russia}
\end{center}

\bll
\hspace{0.8 truecm}
\parbox{14.4 truecm}{
An explanation is suggested for the north-south asymmetry of the polar magnetic field reversal in the current cycle of solar activity. The contribution of the Babcock-Leighton mechanism to the poloidal field generation is estimated using sunspot data for the current activity cycle. Estimations are performed separately for the northern and southern hemispheres. The contribution of the northern hemisphere exceeded considerably that of the southern hemisphere during the initial stage of the cycle. This is the probable reason for the earlier reversal of the northern polar field. The estimated contributions of the Babcock-Leighton mechanism are considerably smaller than similar estimations for the previous activity cycles. A relatively weak (<1\,G) large-scale polar field can be expected for the next activity minimum.
 }

\bll


\newpage

\reversemarginpar

\setlength{\baselineskip}{0.8 truecm}

 \centerline{INTRODUCTION}
 \bl
The large-scale (longitude-averaged) magnetic field of the Sun has two basic components. There is a relatively weak ($\sim 1$~G) poloidal field and a much stronger toroidal field whose strength counts in thousands Gauss as indicated by the field strength in sunspots. The poloidal field, nevertheless, dominates at high latitudes as the toroidal filed has to vanish at the poles for geometrical reasons. Both field components reverse sing from cycle to cycle of solar activity. 11-year cycles of magnetic energy, or sunspot cycles, and 22-year magnetic cycles can, therefore, be distinguished.

Reversals of the polar poloidal field occur near the maxima of sunspot activity. The current 24th solar cycle is distinguished by relatively high north-south asymmetry of polar field reversal: the northern polar field reversed its sign more than one year before the southern field (Svalgaard \& Kamide 2013; Benevolenskaya 2013; Mordvinov \& Yazev 2014). This paper suggests an explanation for this asymmetry from the standpoint of the dynamo theory.

The dynamo theory explains magnetic cycles by two basic effects - generation of a toroidal field from a poloidal one by differential rotation and conversion of the toroidal field back to a poloidal configuration by relatively small-scale cyclonic motions (cf., e.g., the monograph of Parker 1979). These two effects are conventionally named the $\Omega$- and $\alpha$-effects respectively. The dynamo operates deep in the solar convection zone. Observational information about its two basic effects is scarce. Therefore, recently obtained evidence, based on sunspot data, for the action of one of the species of the $\alpha$-effect known as the Babcock-Leighton mechanism on the Sun (Erofeev 2004; Dasi-Espuig et al. 2010; Olemskoy et al. 2013) is valuable. The Babcock-Leightom $\alpha$-effect is probably the principal mechanism for poloidal field generation in the Sun (Choudhuri 2011; Kitchatinov \& Olemskoy 2012). A special feature of this effect is the possibility of estimating its contribution to poloidal field formation from sunspot data. In this paper, estimations are performed using the DPD-catalogue of sunspots (Gy\H{o}ri et al. 2011; Lefevre \& Clette 2014). The estimations are done separately for the northern and southern hemispheres. We shall see that during the growth phase of the current cycle of solar activity, the Babcock-Leighton $\alpha$-effect in the northern hemisphere was considerably larger compared to the southern hemisphere. Most likely, this was the reason for earlier reversal of the northern polar field.
 \bll
 \centerline{THE METHOD}
 \bl
The Babcock-Leighton mechanism is related to Joy's law for sunspot groups: spots leading in rotational motion are located on average closer to the equator than following spots. The leading and following spots usually have magnetic fields of opposite polarities. The tilt angle $\alpha$ of the line connecting the centers of gravity of opposite polarities to the local solar parallel is considered positive if the polarity of the leading spots of a given cycle is located closer to the equator and negative otherwise. Joy's law states that the mean value of the tilt angle is positive and increases with latitude (Hale et al. 1919). Magnetic fields of active regions possess poloidal components, which contribute to the mean poloidal field of the Sun upon the active regions' decay. This mechanism for the poloidal field generation was proposed by Babcock (1961) and first used in the dynamo model by Leighton (1969). Upon the active regions' decay and subsequent diffusion of their magnetic fields over the solar surface, a major part of magnetic flux is lost due to annihilation of opposite polarities. Only a small part ($\sim 0.1$\%) of total magnetic flux of the active regions contributes to the new global poloidal field (Charbonneau 2010).

The following estimation for the contribution of the Babcock-Leighton $\alpha$-effect to the poloidal field was suggested (Kitchatinov \& Olemskoy 2011):
\begin{equation}
    B = \sum\limits_{i}^{} S_i \ell_i \sin\alpha_i ,
    \label{1}
\end{equation}
where summation is over the solar active regions, $S$ is the area of the largest spot of an active region (including penumbra) in millionth parts of the solar hemisphere, $\ell$ is the distance between the centers of gravity of opposite polarities in kilometers and $\alpha$ is the tilt angle. Parameters on the right-hand side of (\ref{1}) are taken at the time of maximum development of a sunspot group. $B$ of equation (\ref{1}) is proportional to the total contribution of the active regions to a large-scale poloidal field. Field strength in sunspots varies in a not too broad range around 3000\,G. The magnetic flux of an active region is therefore assumed proportional to the area $S$. The poloidal part of the active region magnetic field is proportional to the meridional projection of the distance $\ell$, i.e. to the quantity  $\ell\sin\alpha$, and it changes sign if the sign of this quantity reverses. Expansion of the contribution of an active region to the poloidal field in powers of $\ell\sin\alpha$ should, therefore, include only odd powers. Equation (\ref{1}) retains only the first linear term of the expansion in the small parameter $\ell\sin\alpha /R_\odot$.

The $B$-values (\ref{1}) estimated for individual solar cycles correlate well with the polar field in the activity minima following these cycles (Olemskoy et al. 2013). The correlation evidences participation of the Babcock-Leighton $\alpha$-effect in the solar dynamo. In this paper, summation in (\ref{1}) is performed separately for the northern and southern solar hemispheres. The difference between the hemispheric contributions to the poloidal field is used for interpreting the observed north-south asymmetry of polar fields. Parameters of the eqaution (\ref{1}) were taken from the DPD-catalogue of sunspots (http://fenyi.solarobs.unideb.hu/ DPD/index.html; Gy\H{o}ri et al. 2011). Data on the polar magnetic fields were taken from the catalogue of the Wilcox Solar Observatory (http://wso.stanford.edu/Polar.html; Svalgaard et al. 1978).
 \bll
 \centerline{RESULTS AND DISCUSSION}
 \bl
Figure 1 shows the $B$-values for the northern and southern hemispheres as the functions of time. Summation in equation (\ref{1}) was performed by including contributions of the active regions from the last activity minimum to the current date, on which $B$ thus depends. Contributions of both hemispheres to the poloidal field do not steadily increase with time. The contributions decrease sometimes. This is because the distribution of the tilt angles $\alpha$ is rather broad and extends into the region of negative values (see, e.g., Figure 11 in Howard 1996). In other words, considerable fluctuations are inherent to the Babcock-Leighton $\alpha$-effect (Olemskoy et al. 2013). The fluctuations can be caused by random flows in the solar convection zone through which the magnetic fields of active regions rise to the surface, as well as by the superposition of fluctuating and global toroidal fields near the convection zone base from where they rise (Khlystova \& Sokoloff 2009).

\begin{figure}[htb]
 \centerline{
 \includegraphics[width=12 cm]{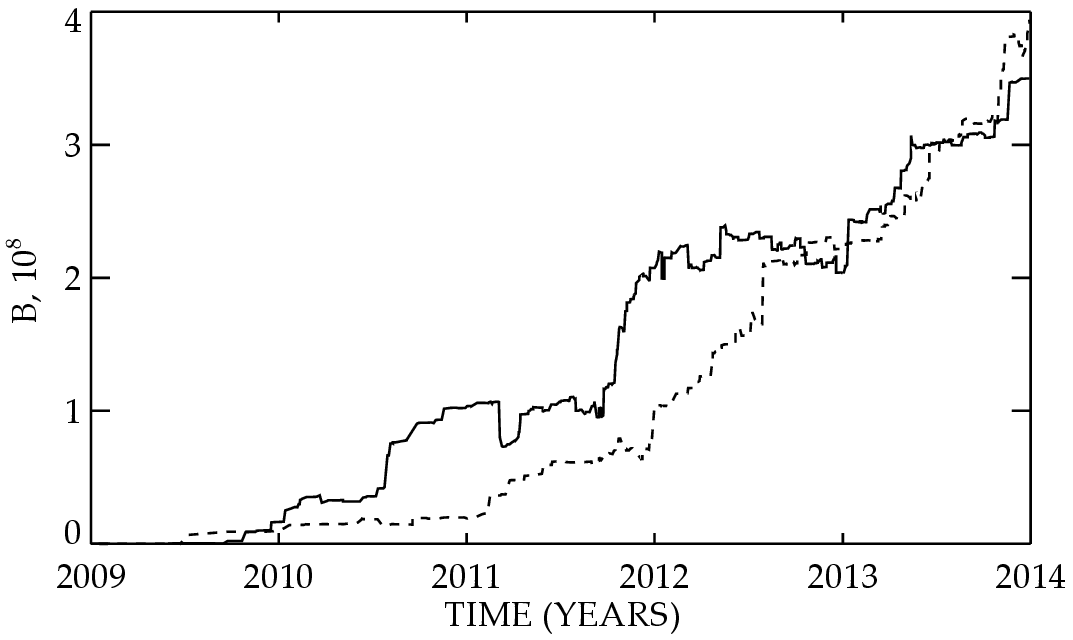}}
 \begin{description}
 \item{\small Fig.~1. The $B$-values of equation (\ref{1}) estimated separately for the northern (full line) and southern (dashed line) hemispheres. The summation (\ref{1}) covers the time interval from the last activity minimum to the current date. The $B$-values are thus date-dependent.
    }
 \end{description}
\end{figure}

Fluctuations are important for the solar dynamo. Variability of the amplitudes and durations of the activity cycles and even the Grand activity minima can be explained by the fluctuations (Choudhuri \& Karak 2012). The fluctuations may also be the reason for north-south asymmetry of magnetic activity (Olemskoy \& Kitchatinov 2013). It can be seen in Fig.1 that the northern hemisphere contributed more to the new poloidal field formation than the southern hemisphere at the beginning of the current activity cycle. Contributions of both hemispheres become almost equal by the end of 2013 and the $B$-value for the southern hemisphere even exceeded that of the northern hemisphere by the beginning of 2014.

\begin{figure}[htb]
 \centerline{
 \includegraphics[width=12 cm]{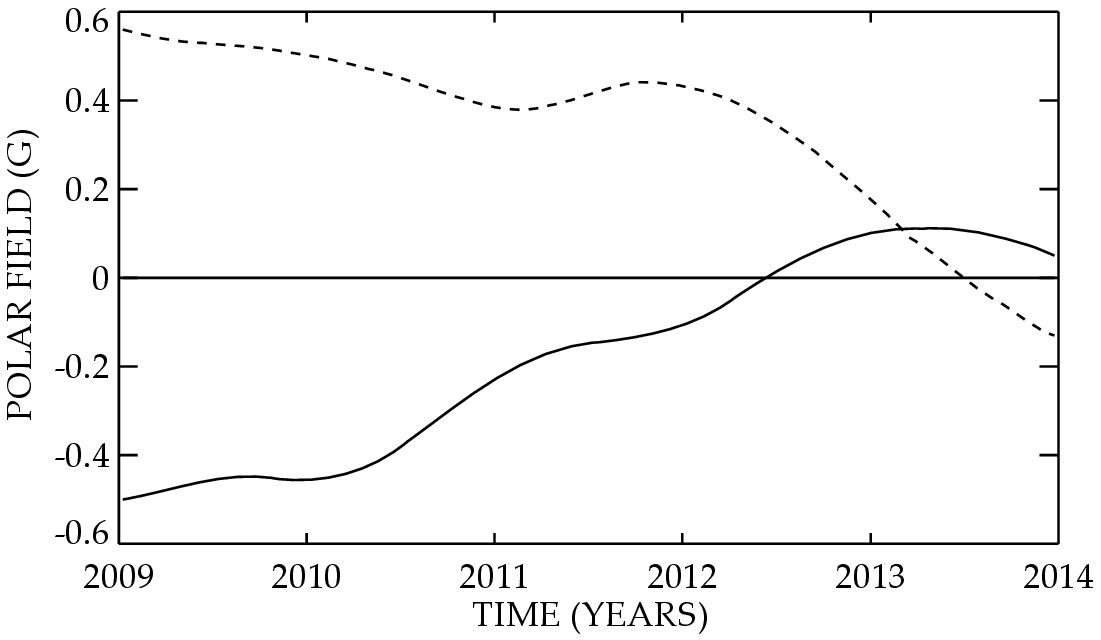}}
 \begin{description}
 \item{\small Fig.~2. Mean magnetic field of the northern (full line) and southern (dashed) polar regions of the Sun in the current activity cycle from the data of the Wilcox Solar Observatory.
    }
 \end{description}
\end{figure}

Figure 2 shows that the northern polar field reversed its sign in the middle of 2012 when the $B$-values of both hemispheres were already nearly equal. The Babcock-Leighton $\alpha$-effect acts where active regions are situated, i.e. at the middle and low latitudes. Accordingly, the poloidal field reversal occurs initially at middle latitudes and it takes more than a year for the \lq new' poloidal field to reach the poles (Mordvinov \& Yazev 2013; 2014). Reversals of the polar fields in Fig.2 are related to the preceding in time $B$-values. The northern hemisphere dominated in these earlier epochs (Fig.1). Earlier reversal of the northern polar field in the current solar cycle is brought about by the \lq higher activity' of the $\alpha$-effect in the northern hemisphere.

The above consideration gives one more evidence for participation of the Babcock-Leighton $\alpha$-effect in the solar dynamo. Other species of the $\alpha$-effect may also participate (Passos et al. 2014) but the Babcock-Leighton $\alpha$-effect, most likely, dominates.

It may be noted in Fig.2 that the northern polar field decreased in the second half of 2013. This is probably related to the decrease in $B$-value of the northern hemisphere in 2012 (Fig.1). At this time, active regions giving negative contributions to the sum of equation (\ref{1}) were obviously present in the northern hemisphere to decrease the contribution of the Babcock-Leighton mechanism to the new poloidal field.

Judging from Fig.1, the total $B$-value for both hemispheres was $\simeq 7\times 10^8$ by the beginning of 2014. This is about four times smaller than the corresponding values of activity cycles 19-21 (Kitchatinov \& Olemskoy 2011). The current activity cycle is not yet completed, but if we do not see a considerable increase in $B$-value in its remainder, the polar field in the coming activity minimum will be weak (<1\,G) similar to the previous minimum. As the strength of activity cycles is proportional to the polar field of preceding minima (Jiang et al. 2007), this in turn may result in the next 25th solar cycle being low.
 \bll

This work was supported by the Russian Foundation for Basic research (Project 13-02-00277).
\bll

\centerline{REFERENCES}
\begin{description}
\item Babcock,~H.\,W.
    1961, \apj\ {\bf 133}, 572
\item Benevolenskaya,~E.\,E.
    2013, Geomagnetism and Aeronomy {\bf 53}, 891
\item Charbonneau,~P.
    2010, Living Rev. Solar Phys. {\bf 7}, 3
\item Choudhuri,~A.\,R.
    2011, in {\sl The Physics of the Sun and Star Spots, IAU Symposium 273}
    (Eds. D.~Choudhary \& K.\,G.~Strassmeier, Cambridge Uni. Press), p.28
\item Choudhuri,~A.\,R., \& Karak,~B.\,B.
    2012, Phys. Rev. Lett. {\bf 109}, 171103
\item Dasi-Espuig,~M., Solanki,~S.\,K., Krivova,~N.\,A. et al.,
    2010, \aa\ {\bf 518}, A7
\item Erofeev,~D.\,V.
    2004, in {\sl Multi-Wavelength Investigations of Solar Activity, IAU Sym\-po\-si\-um 223} (Eds. A.\,V.~Stepanov, E.\,E.~Benevolenskaya, \& A.\,G.~Kosovichev, Cam\-b\-rid\-ge Uni. Press), p.97
\item Gy\H{o}ri,~L., Baranyi,~T., \& Ludm\'any,~A.
    2011, in {\sl The Physics of the Sun and Star Spots, IAU Symposium 273}
    (Eds. D.~Choudhary \& K.\,G.~Strassmeier, Cambridge Uni. Press), p.403
\item Hale,~G.\,E., Ellerman,~F., Nicholson,~S.\,B., \& Joy,~A.\,H.
    1919, \apj\ {\bf 49}, 153
\item Howard,~R.\,F.
    1996, \araa\ {\bf 34}, 75
\item Jiang,~J., Chatterjee,~P., \& Choudhuri,~A.\,R.
    2007, \mnras\ {\bf 381}, 1527
\item Khlystova,~A.\,I., \& Sokoloff,~D.\,D.
    2009, \aj\ {\bf 53}, 281 (2009).
\item Kitchatinov,~L.\,L., \& Olemskoy,~S.\,V.
    2011, \paj\ {\bf 37}, 656
\item Kitchatinov,~L.\,L., \& Olemskoy,~S.\,V.
    2012, \sp\ {\bf 276}, 3
\item Lefevre,~L., \& Clette,~F.
    2014, \sp\ {\bf 289}, 545
\item Leighton,~R.\,B.
    1969, \apj\ {\bf 156}, 1
\item Mordvinov,~A.\,V., \& Yazev,~S.\,A.
    2013, \aj\ {\bf 57}, 448
\item Mordvinov,~A.\,V., \& Yazev,~S.\,A.
    2014, \sp\ {\bf 289}, 1971
\item Olemskoy,~S.\,V., \& Kitchatinov,~L.\,L.
    2013, \apj\ {\bf 777}, 71
\item Olemskoy,~S.\,V., Choudhuri,~A.\,R., \& Kitchatinov,~L.\,L.
    2013, \aj\ {\bf 57}, 458
\item Parker,~E.\,N.
    1979, {\sl Cosmical Magnetic Fields: Their Origin and Their Activity} (Oxford, Clarendon Press)
\item Passos,~D., Nandy,~D., Hazra,~S., \& Lopes,~I.
    2014, \aa\ {\bf 563}, A18
\item Svalgaard,~L., Duvall,~T.\,L.,\,Jr., \& Scherrer,~P.\,H.
    1978, \sp\ {\bf 58}, 225
\item Svalgaard,~L., \& Kamide,~Y.
    2013, \apj\ {\bf 763}, 23
\end{description}
\end{document}